\documentclass[aps,prl,twocolumn,groupedaddress]{revtex4}

\usepackage{amsfonts}
\usepackage{graphicx}
  \usepackage{amsthm}
\usepackage{hyperref}

\begin{document}


\title{Quantum mechanics is the square root of a stochastic process}


\author{Marco Frasca}
\email[]{marcofrasca@mclink.it}
\affiliation{Via Erasmo Gattamelata, 3 \\ 00176 Roma (Italy)}


\date{\today}

\begin{abstract}
We prove a theorem showing that quantum mechanics is not directly a stochastic process characterizing Brownian motion but rather its square root. This implies that a complex-valued stochastic process is involved. Schr\"odinger equation is immediately derived without further assumptions using It\=o integrals that are properly generalized. Fluctuations in space arise from a Brownian motion and the combined effect of a stochastic process with a symmetric Bernoulli distribution typical of tossing a coin.
\end{abstract}


\maketitle


Formal correspondence between heat and Schr\"odinger equations prompted some researchers to try to identify an underlying stochastic process to quantum behavior and that randomness could arise by some classical mechanism. If this would be true, we will be able to reach an understanding of the formalism through some kind of classical behavior where probabilistic behavior could arise from some underlying degrees of freedom. In this area it should be remembered the pioneering work of Edward Nelson \cite{nels} and subsequent deep analysis due to Francesco Guerra and his group \cite{Guerra:1981ie} that termed this reformulation of quantum mechanics as ``stochastic mechanics''. This approach matches directly a Wiener process to the Schr\"odinger equation passing through a Bohm-like set of hydrodynamic equations and so, it recovers all the drawbacks of Bohm formulation. But beside this, this view met severe criticisms motivating some researchers to a substantial claim that ``no classical stochastic process underlies quantum mechanics'' \cite{hang}, showing contradiction with predictions of quantum mechanics, and subsequent attempts to partially or fully recover this view with non-Markovian processes \cite{skor} or repeated measurements \cite{bla1,bla2,bla3}.

The question of a fluctuating space-time is a critical aspect in studies of quantum gravity and the idea that some Brownian motion is underlying this quantum behavior has emerged in literature \cite{Kiefer:1989cv}. So, an understanding of the stochastic nature of quantum mechanics could play a pivotal role in this area of research.

The reason of the difficulties that have been met on this analysis can be traced back to the fact that time in quantum mechanics is just rotated onto an imaginary axis \cite{wick} and this completely changes the very nature of the random process we should consider. Typically, a state that is well-localized in space, evolving in time with the free Schr\"odinger equation, has a probability density with a variance spreading quadratically in time, rather than linearly as happens for the Brownian motion, making clear that we are not coping with a typical Wiener process and something different is happening.

In this letter we will prove a mathematical theorem for stochastic processes recovering in a straightforward way the Schr\"odinger equation as the underlying equation to the diffusion process. The main idea is that what we observe in a quantum mechanical behavior is the square root of a stochastic process rather than a classical stochastic process. This will have the striking result to show that, underlying the weird quantum behavior, there is just a simple stochastic process, a Bernoulli process, typical of tossing a coin. In the following we will largely use It\=o calculus \cite{gard,kleb}. 

In order to prove the main theorem of this letter, we have to evaluate the stochastic process $dX(t)=[dW(t)]^\frac{1}{2}$. For integer powers it is known that \cite{kleb} $[dW(t)]^n=0$ for $n\in\mathbb{N}$ and $n>2$, $[dW(t)]^2=d[W,W](t)=dt$ and finally $dWdt=dtdW=0$ that we will use in our proof. We note that these are just formal definitions of properties of It\=o integral and cannot be considered out of this. Here we extend stochastic calculus to some different processes generally not found in the current literature. In order to accomplish our aims we give the following definitions for the absolute value of a Wiener process

\theoremstyle{definition}
\newtheorem*{defn1}{Absolute value process}
\begin{defn1}
Given the Wiener process $dW$, we decompose it into two processes ${\rm sign}(dW)$ and $|dW|$ sot that $dW={\rm sign}(dW)|dW|$.
\end{defn1}

\theoremstyle{definition}
\newtheorem*{defn2}{Generalized It\=o integral}
\begin{defn2}
A generalized Wiener process $[dW]^\alpha$ with $\alpha\in\mathbb{R}$ and $\alpha>0$ and $|dW|$ must be intended in the sense of the It\=o integral when, given the sums $S_n=\sum_{i=1}^nG(\tau_i)(W(t_i)-W(t_{i-1}))^\alpha$ or $S_n=\sum_{i=1}^nG(\tau_i)|W(t_i)-W(t_{i-1})|$ there exists finite a limit $S$ for $n\rightarrow\infty$, in the root mean square sense, $\lim_{n\rightarrow\infty}\langle(S_n-S)^2\rangle=0$.
\end{defn2}

This definition entails some difficulties as the Riemann sums could not converge. This means that we must properly regularize them. Let us show an example due to Didier Piau \cite{piau}: We want to evaluate the simplest integral $\int_{t_0}^t|dW|$. We have to evaluate $\langle S_n^2\rangle$ in the limit $n\rightarrow\infty$. But we have $\langle|W(t+s)-W(t)|\rangle=\sqrt{2s/\pi}$ for $s,\ t>0$ and one should remeber that increments are independent so, $\langle|W(t_i)-W(t_{i-1})||W(t_k)-W(t_{k-1})|\rangle=(2/\pi)\sqrt{t_i-t_{i-1}}\sqrt{t_k-t_{k-1}}$. Then, choosing $t_i=i/n$ implies the limit for $n\rightarrow\infty$ of $(2/\pi)(1/n)(\sum_{i=1}^n)^2$ that, if we do not regularize it properly, runs too infinity. In order to give a regularization technique, we refer to \cite{hardy}. So, we set $\sum_{i=1}^\infty=\zeta(0)=-1/2$ and $\sum_{i=1}^\infty(-1)^n=-1/2$. This implies that Riemann sums must be always intended after a proper regularization. From a similar argument we get that ${\rm sign}(dW)$ has null measures as It\=o integral while its product with $|dW|$ has not being an ordinary Wiener process, provided we consider $\int_{t_0}^t{\rm sign}(dW)$ defined through the sums $S_n=\sum_{i=1}^n{\rm sign}(W(t_i)-W(t_{i-1}))$ and the limit $n\rightarrow\infty$ in the root mean square sense with the given regularization. Finally, we note that $(dW)^\alpha=0$ for $\alpha>2$ \cite{lowt}.

This class of It\=o integrals is interesting because in most cases, as we will show, are computable in the standard form $adW(t)+bdt$ and so this limit has a well defined meaning. In order to give an example, let us consider the case $\alpha=2$. It is easy to verify that one has that
\begin{equation}
   \langle S_n\rangle=\sum_iG(\tau_i)\langle(W(t_i)-W(t_{i-1}))^2\rangle=\sum_iG(\tau_i)(t_i-t_{i-1})
\end{equation}
that gives the standard limit $\int_{t_0}^tdt'G(t')$ as it should. Indeed, this is the known case formally written in stochastic calculus as $(dW)^2=dt$. So, this idea is generalized given the above definition. But in order to complete this definition we consider also null measure processes, in the It\=o sense given above, like ${\rm sign}(dW(t))$ that we will discuss below. Similarly, we expect that integration with respect to a process like $a+b{\rm sign}(dW(t))$ will produce a constant as we moved the mean from 0 to a non-null finite value, after regularization. In this way we can generalize a stochastic process to a form $\mu_0\Phi(t)+\mu_1dW+\mu_2dt$ having $\Phi(t)$ a finite constant value, eventually zero, after integration. It important to emphasize that the change of signs in a Wiener increment is regulated by a symmetric Bernoulli distribution with $p=1/2$ \cite{Olek}, a fact that we will use below.

In the general case we can state the following:

\newtheorem*{thm1}{Stochastic square root theorem}
\begin{thm1}
Stochastic differential equation $[dW(t)]^\frac{1}{2}$ has the generic expansion $(\mu_0+\mu_1 |dW|+\mu_2 dt)\Phi_\frac{1}{2}(t)$ being $\mu_0$, $\mu_1$ and $\mu_2$ some constants and $\Phi_\frac{1}{2}(t)$ a Bernoulli process taking the values $1$ and $i$ with probability $1/2$ that has the form $\Phi_\frac{1}{2}(t)=\frac{1-i}{2}{\rm sign}(dW(t))+\frac{1+i}{2}$.
\end{thm1}

\begin{proof}
Let us consider the stochastic differential equation $dX(t)=[dW(t)]^\frac{1}{2}$. We reformulate this in the form
\begin{equation}
   dX(t)=[\mu_0+\mu_1 |dW(t)|+\mu_2 dt]\Phi_\frac{1}{2}(t).
\end{equation}
with $\Phi_\frac{1}{2}(t)$ that will depend on sign changes in the increments of the Wiener process. Now, we show that $[dX(t)]^2=dW(t)$ modulo a null measure process in the It\=o sense. So,
\begin{equation}
   dX(t)^2=[\mu_0+\mu_1 |dW(t)|+\mu_2 dt]^2{\rm sign}(dW(t))
\end{equation}
that gives
\begin{equation}
   [dX(t)]^2=\mu_0^2{\rm sign}(dW(t))+dW(t)
\end{equation}
using techniques of stochastic calculus given in \cite{kleb}, provided we take $\mu_1=1/2\mu_0$, $\mu_2=-1/8\mu_0^3$ and being ${\rm sign}(dW(t))|dW(t)|=dW(t)$ by definition. The Bernoulli process ${\rm sign}(dW(t))$ has a null It\=o integral and so, it represents a null measure process.
\end{proof}

We note that this proof can be iterated with any power of $1/n$ with $n\in\mathbb{N}$. Bernoulli process we extracted with the square root of the Wiener process is just equivalent to toss a coin. Then, we can state the following theorem:

\newtheorem*{thm}{Square root theorem}
\begin{thm}
The square root of a stochastic process implies a diffusional process evolving with the Schr\"odinger equation.
\end{thm}

\begin{proof}
Let us consider the stochastic differential equation
\begin{equation}
    dX(t)=[dW(t)+\beta dt]^\frac{1}{2}.
\end{equation}    
This process cannot be real-valued as the increments of the Wiener process can be negative. We assume $W(t)$ a real-valued stochastic process. Then, using the preceding theorem we can write 
\begin{equation}
\label{eq:dq}
    dX(t)=\left[\frac{1}{2}+|dW(t)|+(-1+\beta{\rm sign}(dW(t)))dt\right]\Phi_{\frac{1}{2}}(t).
\end{equation}
For the following we just note that the Bernoulli process has mean $(1+i)/2$ and variance $-i/2$. We obtained a stochastic differential equation that is indeed a complex-valued stochastic differential equation. We also note that the integral of ${\rm sign}(dW(t))$ is zero and so, the integral of $\Phi_{\frac{1}{2}}(t)$ is just a finite constant.


From this process we can read off the mean and the variance that will be obtained through mean and variance of the process ${\rm sign}(dW(t))$. This gives
\begin{eqnarray}
   \mu &=&-\frac{1+i}{2}+\beta\frac{1-i}{2} \nonumber \\
   \sigma^2 &=&-\frac{i}{4}.
\end{eqnarray}

Then, Kolmogorov equation takes the simplest form
\begin{equation}
   \frac{\partial\psi}{\partial t}=\left(-\frac{1+i}{4}+\beta\frac{1-i}{2}\right)\frac{\partial\psi}{\partial X}-\frac{i}{4}\frac{\partial^2\psi}{\partial X^2}
\end{equation}
representing the Schr\"odinger equation for a free particle after properly fixing $\beta$. 
\end{proof}
 

A first application of this theorem can be seen in the following proposition \cite{fafs} that displays the square-root link between quantum mechanics and stochastic processes:

\newtheorem*{prop}{Farina-Frasca-Sedehi~proposition}
\begin{prop}
For a well-localized free particle there exists a map between the wave function and the square root of binomial coefficients.
\end{prop}

In this case one consider an initial state given by $\psi(x,0)=e^{-\frac{x^2}{2(\Delta x)^2}}/{(\pi\Delta x^2)^\frac{1}{4}}$ that is evolved to $\psi(x,t)$ using the free Schr\"odinger equation. Then, one can define a discrete map that takes this solution to a discrete one $\psi(n,k)$ showing that $|\psi(n,k)|^2$ can be mapped on a classical discrete stochastic process being the distribution of $n$ successes on a row of tossing a coin after division by $2^n$ entering into the map. The result is perfectly consistent with the theorem proved above. Indeed, one can understand this proposition if a random walk is taken for the probability of a well-localized particle on a discrete quantum world. In this case, assuming a symmetric distribution, we can immediately write down
\begin{equation}
   P(n,k)=\left(\begin{array}{c}
    n \\
    k
\end{array}\right)\frac{1}{2^n}
\end{equation}
but the quantum behavior can be extracted from this by a square root of the modulus of a complex distribution as shown in the square root theorem and so the wave function is
\begin{equation}
   \psi(n,k)=\frac{1}{2^\frac{n}{2}}\left(\begin{array}{c}
    n \\
    k
\end{array}\right)^\frac{1}{2}e^{i\theta(n,k)}
\end{equation}
that is the content of the Farina-Frasca-Sedehi proposition. Here, the $2^n$ factor is moved into the wave function with the scalar product defined as $\langle\phi_1,\phi_2\ \rangle=\sum_{k=0}^n\phi_1^*(n,k)\phi_2(n,k)$ that gives $\langle\psi,\psi\rangle=1$. One can work the other way round and put the $2^n$ factor into the definition of the scalar product. Finally, to understand the lattice structure and the phase we need to solve the Schr\"odinger equation as shown in \cite{fafs}.

The next step is the introduction of a potential. This is easily accomplished by generalizing the stochastic process in eq.(\ref{eq:dq}) considering $dV(X,t)=U(X)dX$ a stochastic pre-potential and $U(X)$ the potential acting on the quantum particle. Here we want to obtain the transition probability from the square root theorem in order to show that the claim in \cite{hang} can be evaded. Let us state the question put forward by Grabert, H\"anggi and Talkner (GHT) in \cite{hang}. The argument is straightforward: If quantum mechanics is described by a classical stochastic process, transition probabilities are given by a master equation through a Fokker-Planck operator but this fails to grant the exact values of all the momenta of observables. In our case the situation is largely different as the Fokker-Planck operator turns to be the Schr\"odinger one, at least for a free particle. In order to give an answer to GHT argument we have to show that a potential can be introduced into our stochastic process that, as already pointed out, is not strictly a classical one but its square root. We prove the following corollary

\newtheorem*{cor}{Square root corollary}
\begin{cor}
The square root of a stochastic process implies a diffusional process evolving with the Schr\"odinger equation also for an interacting particle.
\end{cor}

\begin{proof}
Let us consider the stochastic process
\begin{equation}
    dX(t)=[dW(t)+V(X,t)dt]^\frac{1}{2}.
\end{equation}
It is not difficult to see that we have
\begin{eqnarray}
    dX(t)&=&\left\{\frac{1}{2}+|dW|+\right. \nonumber \\
    &&\left.(-1+V(X,t){\rm sign}(dW(t)))dt\right\}\Phi_{\frac{1}{2}}(t)
\end{eqnarray}
and this gives
\begin{equation}
   [dX(t)]^2=\frac{1}{4}{\rm sign}(dW(t))+dW+V(X,t)dt
\end{equation}
again modulo a Bernoulli process having null measure. But this implies immediately that the diffusional process is now
\begin{equation}
   \frac{\partial\psi}{\partial t}=-\frac{1+i}{4}\frac{\partial\psi}{\partial X}
   +\frac{1-i}{4}\frac{\partial}{\partial X}[V(X,t)\psi]-\frac{i}{4}\frac{\partial^2\psi}{\partial X^2}.
\end{equation}
\end{proof}

This corollary removes GHT objection and Schr\"odinger equation is just the consequence of a stochastic Wiener process multiplied by a Bernoulli process provided a square root is taken.

The main result of this letter is that quantum mechanics represents the square root of a standard Wiener process. Such a square root process is characterized by the presence of a Bernoulli process typical of tossing a coin. This is the underlying elementary process besides a standard Brownian motion. But this Bernoulli process is the one that makes time rotate and change the physical nature of the diffusion process. This kind of mathematical behavior can be easily escaped in literature being not too much intuitive and so, the connection between stochastic processes and quantum mechanics appears absolutely not trivial. This entails a new formulation of quantum mechanics.

Our main results shows that the wave function is mathematically equivalent to a standard probability distribution but now we are working with a complex function and this makes the interpretation not so straightforward.

It is also important to point out that here we are showing that the square root of Brownian fluctuations of space are responsible for the peculiar behavior observed at quantum level. This kind of stochastic process is a square root of a Brownian motion that boils down to the product of two stochastic processes: A Wiener process and a Bernoulli process proper to a tossing of a coin. This aspect could be relevant for quantum gravity studies where emergent space-time could be understood once this behavior will be identified in the current scenarios.

As an aside we note that It\=o integral can be generalized to some exotic situations as these theorems are showing. We hope to push further these analyses in view of a deeper understanding of formalism of quantum mechanics. What appears really striking is that, behind the weirdness of the quantum world, is hidden the tossing of a coin. Finally, we note as the underlying structure is inherently probabilistic and so this appears rather as another mathematical way to view quantum systems.

I would like to thank very much Alfonso Farina that proposed a very successful collaboration and Matteo Sedehi for fruitful discussions during this collaboration, both working at Selex Sistemi Integrati. An important help came from Oleksandr Pavlyk at Wolfram Research and I would like to thank him for this. Some mathematicians at \href{http://math.stackexchange.com/}{Mathematics Stackexchange} were instrumental to understand problems with these questions of stochastic processes. Finally, I would like to thank George Lowther for some very helpful comments.

\end{document}